\documentclass[american]{article}
\usepackage[LGR,T1]{fontenc}
\usepackage[latin9]{inputenc}
\usepackage{geometry}
\usepackage{amsmath}
\usepackage{graphicx}
\usepackage{setspace}
\makeatletter

\DeclareRobustCommand{\greektext}{%
  \fontencoding{LGR}\selectfont\def\encodingdefault{LGR}}
\DeclareRobustCommand{\textgreek}[1]{\leavevmode{\greektext #1}}
\ProvideTextCommand{\~}{LGR}[1]{\char126#1}

\providecommand{\tabularnewline}{\\}

\newcommand{\lyxaddress}[1]{
\par {\raggedright #1
\vspace{1.4em}
\noindent\par}
}

\@ifundefined{date}{}{\date{}}
\makeatother

\usepackage{babel}
\begin{document}

\title{Biophysical inference of epistasis and the effects of mutations on
protein stability and function}

\author{Jakub Otwinowski}
\maketitle

\lyxaddress{University of Pennsylvania, Biology Department, jakubo@sas.upenn.edu}
\begin{abstract}
Understanding the relationship between protein sequence, function,
and stability is a fundamental problem in biology. While high-throughput
methods have produced large numbers of sequence-function pairs, functional
assays do not distinguish whether mutations directly affect function
or are destabilizing the protein. Here, we introduce a statistical
method to infer the underlying biophysics from a high-throughput binding
assay by combining information from many mutated variants. We fit
a thermodynamic model describing the bound, unbound, and unfolded
states to high quality data of protein G domain B1 binding to IgG-Fc.
We infer an energy landscape with distinct folding and binding energies
for each substitution providing a detailed view of how mutations affect
binding and stability across the protein. We accurately infer folding
energy of each variant in physical units, validated by independent
data, whereas previous high-throughput methods could only measure
indirect changes in stability. While we assume an additive sequence-energy
relationship, the binding fraction is epistatic due its non-linear
relation to energy. Despite having no epistasis in energy, our model
explains much of the observed epistasis in binding fraction, with
the remaining epistasis identifying conformationally dynamic regions.
\end{abstract}

\subsection*{Author Summary}

Determining how mutations impact protein stability and function is
instrumental in understanding how proteins carry out their biological
tasks, how they evolve, and how to engineer novel proteins. However
measuring differences in function between mutated variants does not
distinguish whether mutations are directly affecting function or are
destabilizing the protein. Here, we fit a thermodynamic model to data
describing how thousands of variants of a protein bind to an antibody
fragment. We accurately infer separate folding and binding energies
in physical units, providing a detailed energy landscape describing
how mutations affect binding and stability across the protein, and
our non-linear model reproduces many of the observed interactions
between sites.

\section*{Introduction}

Deep mutational scanning (DMS) studies have produced detailed maps
of how proteins and regulatory sequences are related to function by
assaying up to millions of mutated variants, and has had many applications,
from identifying viral epitopes to protein engineering \cite{fowler_deep_2014,wrenbeck_deep_2017}.
While these studies aim to understand molecular function and evolution
by collecting large numbers of sequence-function pairs, the full sequence-function
map is very difficult to determine due to the enormity of sequence
space. Different sites in a sequence may not contribute to molecular
function independently, and the effect of a substitution at one site
may depend on the genetic background. This non-additivity, or epistasis,
means that the entire space of possible sequences may have to be explored
to understand molecular function. 

Given the limited data, mathematical modeling is necessary to make
any progress. However, purely statistical inferences are difficult
to interpret in terms of known biology, and can be too flexible to
make reliable predictions \cite{otwinowski_inferring_2014,plessis_how_2016}.
On the other hand, biophysical systems are not arbitrarily complex
as they follow physical laws and structural constraints. In other
words, there is hope that biophysical knowledge can help explain sequence-function
relationships. A powerful assumption is that in between sequence and
function lie relevant intermediate phenotypes for which we can derive
relatively simple relations to sequence and function \cite{bershtein_bridging_2017}.

The stability of a protein's fold is a fundamental molecular phenotype
under selection. Studying how mutations affect stability is a fundamental
challenge in protein science \cite{magliery_protein_2011}, and is
the aim of some DMS studies \cite{araya_fundamental_2012,traxlmayr_construction_2012,kim_high-throughput_2013,olson_comprehensive_2014}.
However, assaying molecular function, such as binding to a ligand,
is a necessary and insufficient measure of stability, in that most
proteins must be folded to function, but do not necessarily function
if folded. In addition, high-throughput techniques, such as proteolysis
assays \cite{rocklin_global_2017}, do not measure free energy, but
measure scores that are correlated with stability. Similarly, scores
from high-throughput binding assays do not measure binding energy
in physical units, and do not distinguish whether changes seen in
variants are due to changes in the overall fold stability or stability
of the binding interface. 

While high-throughput assays often confound function and stability,
these can be separated with a thermodynamic approach. Thermodynamic
approaches have been at the heart of biophysical models applied to
data to quantify the evolution of regulatory sequences \cite{mustonen_evolutionary_2005,mustonen_energy-dependent_2008,kinney_using_2010,lagator_mechanistic_2017},
and proteins \cite{bloom_thermodynamic_2005,wylie_biophysical_2011,echave_biophysical_2017}.
In the context of proteins, there are typically a few relevant conformational
states, and the kinetics are fast enough to reach thermal equilibrium,
with a free energy determining the probability of each state. At a
minimum, a protein has folded and unfolded states, and other states
may be due to binding, mis-folding, or some other conformational changes.
Two-state models have been important in understanding observed patterns
of substitutions in protein evolution \cite{starr_epistasis_2016,bastolla_what_2017,liberles_interface_2012},
and in general, the ensemble of protein conformations generates epistasis
that makes protein evolution difficult to predict \cite{sailer_molecular_2017}.
A powerful simplification is to approximate the total energy by a
sum over site-specific energies (additivity), which has been observed
in most changes to fold stability \cite{wells_additivity_1990,sandberg_engineering_1993}.
However, even with additivity in energy, the probability of a protein
being in a particular state is non-linear with respect to energy and
therefore epistatic with regard to sequence.

In this work, we infer a thermodynamic model that separates folding
and stability in a small bacterial protein, protein G domain B1 (GB1),
a model system of folding and stability, where a recent high-throughput
assay of functional binding to an immunoglobulin fragment (IgG-Fc)
described the epistasis between nearly all pairs of residues \cite{olson_comprehensive_2014}.
We infer a thermodynamic model with two states, bounded, and unbound,
and another model with three states: bound-folded, unbound-folded
and unfolded. The approximation of additivity in energy allows us
to separate how mutations destabilize the binding interface and how
they destabilize the overall fold. We validate these approximations
by predicting independently measured changes in fold stability. We
describe the folding and stability landscape of the protein, identify
which sites contribute most to binding, and explain much of the observed
epistasis without assuming any energetic interactions.

\section*{Results}

\subsection*{\textit{In vitro} selection of protein variants}

Olson et al.~\cite{olson_comprehensive_2014} mutagenized GB1 to
create a library of protein variants which contained all single amino
acid substitutions (1045 variants) and nearly all double substitutions
(536k variants) of a reference or wild-type sequence. The library
was sequenced before and after an\textit{ in vitro }selection assay,
and the fraction of bound protein to IgG-Fc for a variant with sequence
$\sigma$ is $p^{\prime}(\sigma)=\frac{n_{1}(\sigma)}{n_{0}(\sigma)r}$,
where $n_{0}(\sigma)$ and $n_{1}(\sigma)$ are the sequence counts
before and after selection, and $r$ is a global factor that accounts
for systematic differences between initial and final sequencing (see
Methods for a maximum likelihood derivation of $p^{\prime}$). For
convenience, we define \emph{fitness} as the logarithm of the binding
fraction normalized by the wild-type $\sigma^{W}$
\begin{equation}
f(\sigma)=\log\left(\frac{n_{1}(\sigma)}{n_{0}(\sigma)}\frac{n_{0}(\sigma^{W})}{n_{1}(\sigma^{W})}\right),\label{eq:fitness}
\end{equation}
as an analogy to the growth rate of an exponentially growing population,
although we do not imply that this is the (relative) growth rate of
an organism with this variant.

With nearly every possible double substitution it is possible to study
interactions between sites, or epistasis. We define pairwise epistasis
in fitness as the difference between the fitness of the double mutants
relative to the wild-type and the expectation of additivity, i.e.,
the sum of the fitness of two single mutants:
\begin{equation}
J_{ij}^{ab}=f(\sigma_{/(i,a)/(j,b)}^{W})-f(\sigma_{/(i,a)}^{W})-f(\sigma_{/(j,b)}^{W}).\label{eq:epistasis}
\end{equation}
 where $/(i,a)$ and $/(j,b)$ indicate substitutions at positions
$i,j$ with amino acids $a,b$.

While changes in fitness across all single and double mutants show
where a protein is sensitive to binding, such changes are not informative
of whether mutations are destabilizing the binding interface or the
overall fold, as changes in either one influence the fraction of bound
protein. A thermodynamic model is necessary to separate these effects.

\subsection*{Thermodynamic models}

\begin{figure}
\begin{centering}
\includegraphics[scale=0.6]{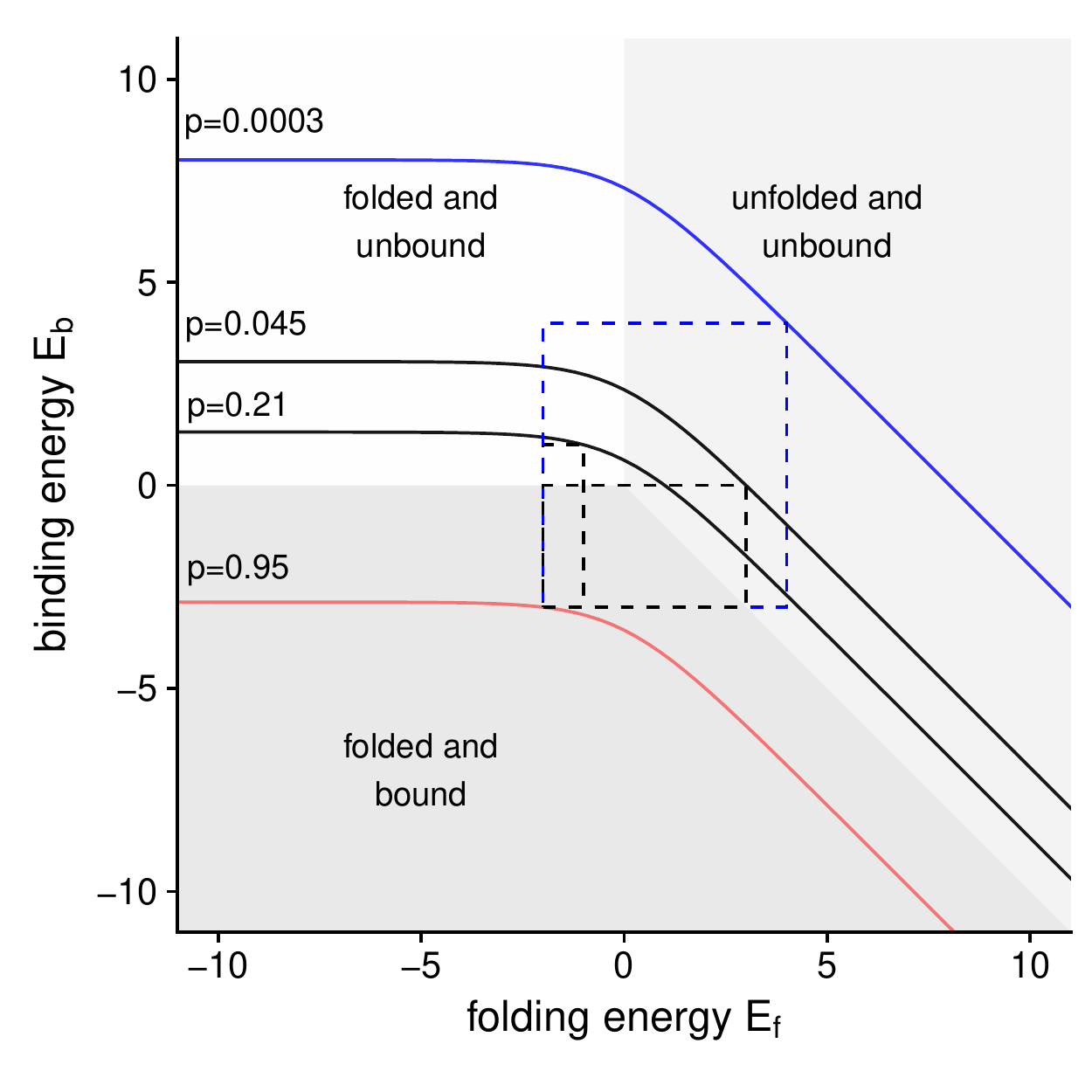}
\par\end{centering}
\caption{\label{fig:ped}Thermodynamic model of three protein states: unfolded
and unbound, folded and unbound, and folded and bound, described by
eq.~\ref{eq:p}. Shaded areas correspond to regions in energy space
where the labeled state is dominant. Given sequences and binding fractions
$p$, the non-linear Boltzmann form (eq.~\ref{eq:p}) imposes constraints
on the possible parameters (additive energies). Solid lines are the
energies compatible with $p$ for four hypothetical sequences: wild-type
(red), two single mutants (black) and a double mutant (blue). Dashed
lines represent additive energies, and connect the wild-type to the
single mutants (black) and the double mutant (blue), which has lengths
equal to the sum of additive energies.}

\end{figure}

Proteins fold into complicated structures and interact with other
molecules depending on the free energy of their different states or
conformations. Under natural conditions, protein states reach thermodynamic
equilibrium very quickly and the Boltzmann distribution relates the
probability of state $i$ to the free energy $E_{i}$: $p_{i}(\sigma)=\frac{1}{Z}e^{-E_{i}(\sigma)}$,
where $E_{i}(\sigma)$ are in dimensionless units and $Z$ is the
normalization factor over states \cite{phillips_physical_2012}. For
a two state bound/unbound model, the fraction of bound protein is
$\frac{1}{1+e^{E(\sigma)}}$, with energy $E(\sigma)$. In order to
separate binding and stability, we define three states: unfolded and
unbound, folded and unbound, and folded and bound, and therefore the
fraction of bound protein is
\begin{equation}
p(\sigma)=\frac{e^{-E_{f}(\sigma)-E_{b}(\sigma)}}{1+e^{-E_{f}(\sigma)}+e^{-E_{f}(\sigma)-E_{b}(\sigma)}}=\frac{1}{1+e^{E_{b}(\sigma)}(1+e^{E_{f}(\sigma)})}\label{eq:p}
\end{equation}
with folding energy $E_{f}(\sigma)$ and binding energy $E_{b}(\sigma)$.
The folding energy is relative to the unfolded state, whereas the
binding energy is relative to the folded-unbound state, up to a constant
that depends on the concentration of ligand (the chemical potential).
Importantly, this binding energy measures only the destabilization
of the binding interface, and is distinct from dissociation constants
that are related to our model by $K_{d}\propto e^{E_{b}}(1+e^{E_{f}})$
(neglecting the chemical potential). Intuitively, low binding and
folding energy leads to large $p$, and the shaded areas in Fig.~\ref{fig:ped}
show regions in energy space where each label indicates the most likely
state.

Given an experimentally measured $p$ of a single variant, there are
many values $E_{b}$ and $E_{f}$ that match $p$, and it is not possible
to identify these energies, as shown by the contour lines of equal
$p$ in Fig.~\ref{fig:ped}. However, with the approximation of additivity
in energy over sites, it is possible to combine information from many
sequences to estimate energies. Additivity means that the energy is
a sum over energies specific to each substitution $\epsilon(i,a)$:
\begin{align}
E_{f}(\sigma) & =\epsilon_{f}^{W}+\sum_{i}\epsilon_{f}(i,\sigma_{i})\label{eq:ef}\\
E_{b}(\sigma) & =\epsilon_{b}^{W}+\sum_{i}\epsilon_{b}(i,\sigma_{i})\label{eq:eb}
\end{align}
where subscripts $f$ and $b$ indicate folding and binding respectively,
$\sigma_{i}$ is the amino acid at position $i$, and $\epsilon(i,\sigma_{i}^{W})=0$
so that the wild-type energy is $\epsilon^{W}$. While additivity
has been observed in many experimental measurements of changes in
fold stability \cite{wells_additivity_1990,sandberg_engineering_1993},
it is a local approximation that is likely to be violated for highly
mutated sequences.

To illustrate how multiple data points constrain this non-linear model,
consider a hypothetical quartet of sequences and their measured binding
fraction $p$: the wild-type, two sequences with single substitutions,
and a sequence with both of those substitutions. The lines in Fig.~\ref{fig:ped}
are the energy coordinates consistent with the given $p$, and the
dashed lines are the additive energies that connect the wild-type
(red line) to the single mutants (black lines), and to the double
mutant (blue line). The parameters are not constrained given a wild-type
$p$ and single mutant $p$, as a rectangles can be placed anywhere
between two lines as long as the opposing corners land on them. However,
when considering all four sequences the largest rectangle must have
lengths that are the sum of the smaller rectangle lengths due to additivity,
and the non-linearity of the curves constrains the parameters (additive
energies) that can fit the data, with more data providing more constraints.

We use all sequences and associated counts in a maximum a likelihood
framework to infer all additive and wild-type folding and binding
energies, converted to kcal/mol (see Methods). For comparison, we
also infer energies of the two state model. In Methods, we modify
the likelihood to account for non-specific background binding, and
describe a procedure to overcome local optima via bootstrapping. 

\subsection*{Inferred energy landscape}

\begin{figure}
\begin{centering}
\includegraphics[scale=0.6]{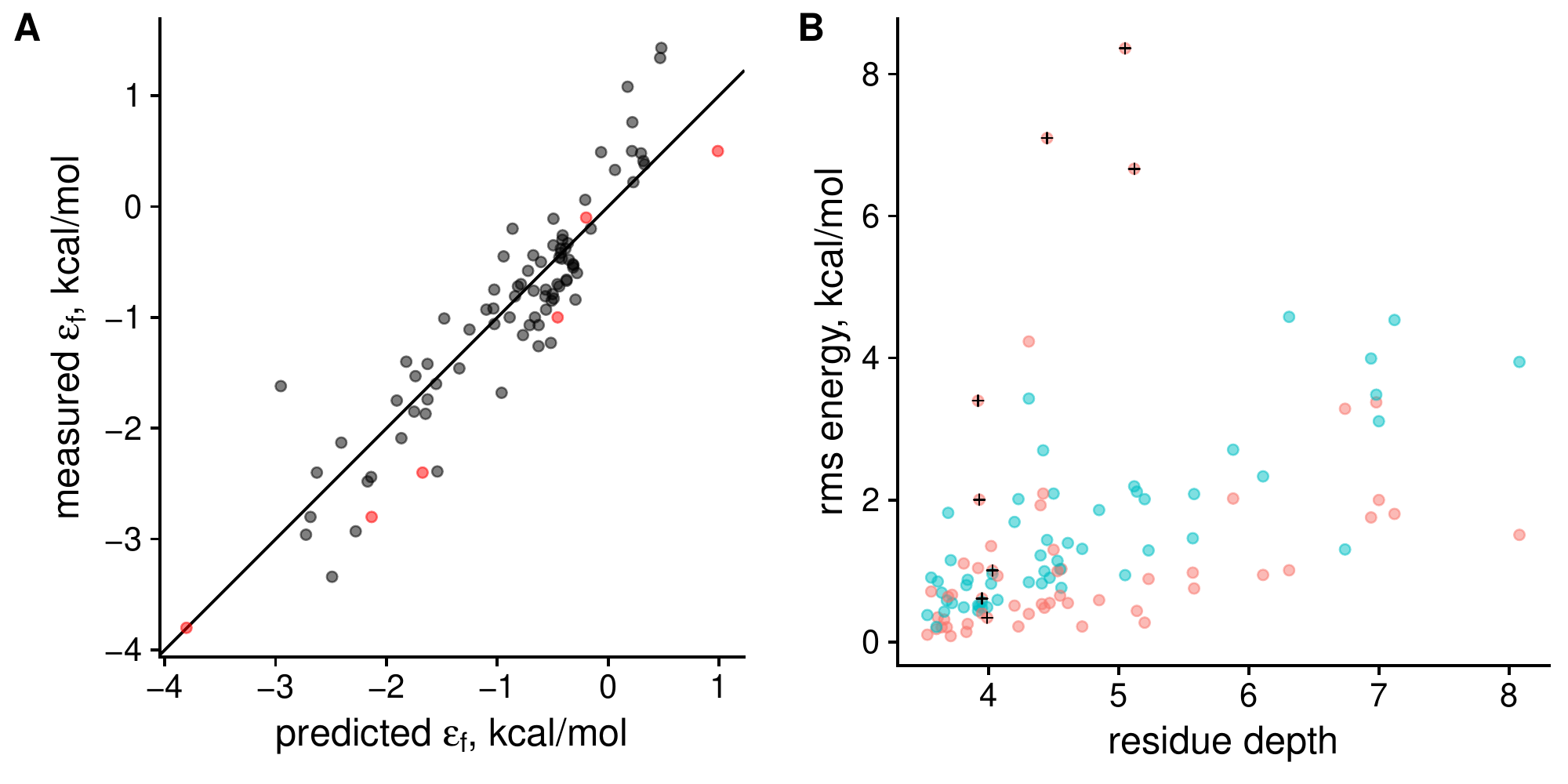}
\par\end{centering}
\caption{\label{fig:prediction} A) Accurate prediction of changes in folding
energy $\epsilon_{f}$ (eq.~\ref{eq:ef}, commonly referred to as
$\Delta\Delta G$) by fitting a three state thermodynamic model to
deep mutational scanning data. Predicted energies have a root mean
square error of 0.39 kcal/mol and $\rho=0.91$ compared to independent
measurements of $\epsilon_{f}$ for 81 single substitutions \cite{olson_comprehensive_2014}.
Six variants with 2-6 mutations are shown in red. The line has a slope
of unity. B) Folding energies (teal) have a stronger relation to residue
depth than binding energies (red). Root mean square energy changes
at each position are shown, and a plus sign indicates sites at the
protein-protein interface \cite{sauer-eriksson_crystal_1995,sloan_dissection_1999}). }
\end{figure}

We compare the inferred additive folding energies to independent low-throughput
measurements of 81 single substitutions in Fig.~\ref{fig:prediction}A
(collected from different sources in \cite{olson_comprehensive_2014}).
The three state model (bound/unbound/unfolded) accurately predicts
$\epsilon_{f}$ in physical units with an root mean squared error
of 0.39 kcal/mol and a correlation of $\rho=0.91$, which is better
than computational methods (\textasciitilde{}0.6 to \textasciitilde{}0.7)
and close to the amount of correlation between replicates of low-throughput
methods (\textasciitilde{}0.86) \cite{potapov_assessing_2009}. Six
variants with 2-6 mutations are also predicted (Fig.~\ref{fig:prediction}A
red) with comparable accuracy. However, 2 highly stable variants (G\textgreek{b}1-c3b4
with 7 mutations, and M2 with 4, not shown) are underestimated by
2.1 and 5.3 kcal/M respectively, suggesting the presence of significant
synergistic epistasis in the folding energy for these variants. 

The two state model (bound/unbound) fits the data similarly well to
the three state model, with a correlation between predicted fitness
$\hat{f}(\sigma)=\log(p(\sigma)/p(\sigma^{W}))$ and measured fitness
of 96.4\% and 97.1\% for two and three state models respectively (see
Fig.~\ref{fig:yyhat}). However, the additive energies inferred by
the two state model have no relation to the independently measured
$\epsilon_{f}$ (Fig.~\ref{fig:ddg1}). Clearly a three state model
is necessary to predict folding energy, and has the added benefit
of estimating binding energy. 

\begin{figure}
\begin{centering}
\includegraphics[scale=0.5]{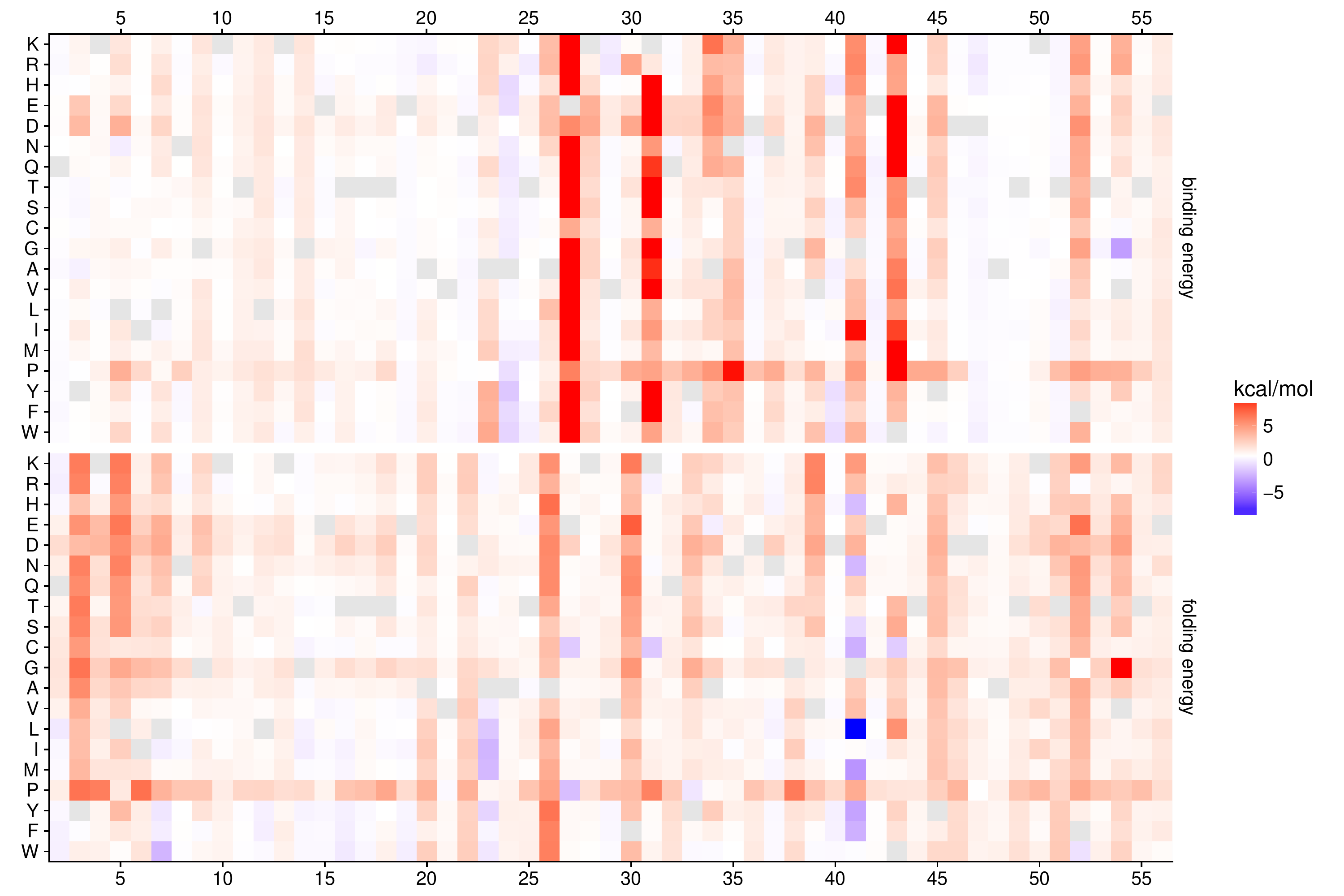}
\par\end{centering}
\caption{\label{fig:efiebi}Inferred additive binding $\epsilon_{b}$ and folding
$\epsilon_{f}$ energies show strikingly different patterns. Three
of the binding sites (27, 31, 43) have strong effects on binding.
Many substitutions at positions 23 and 41 are beneficial for folding
and deleterious for binding, although overall substitutions are uncorrelated. }
\end{figure}

\begin{figure}
\begin{centering}
\includegraphics[width=1\textwidth]{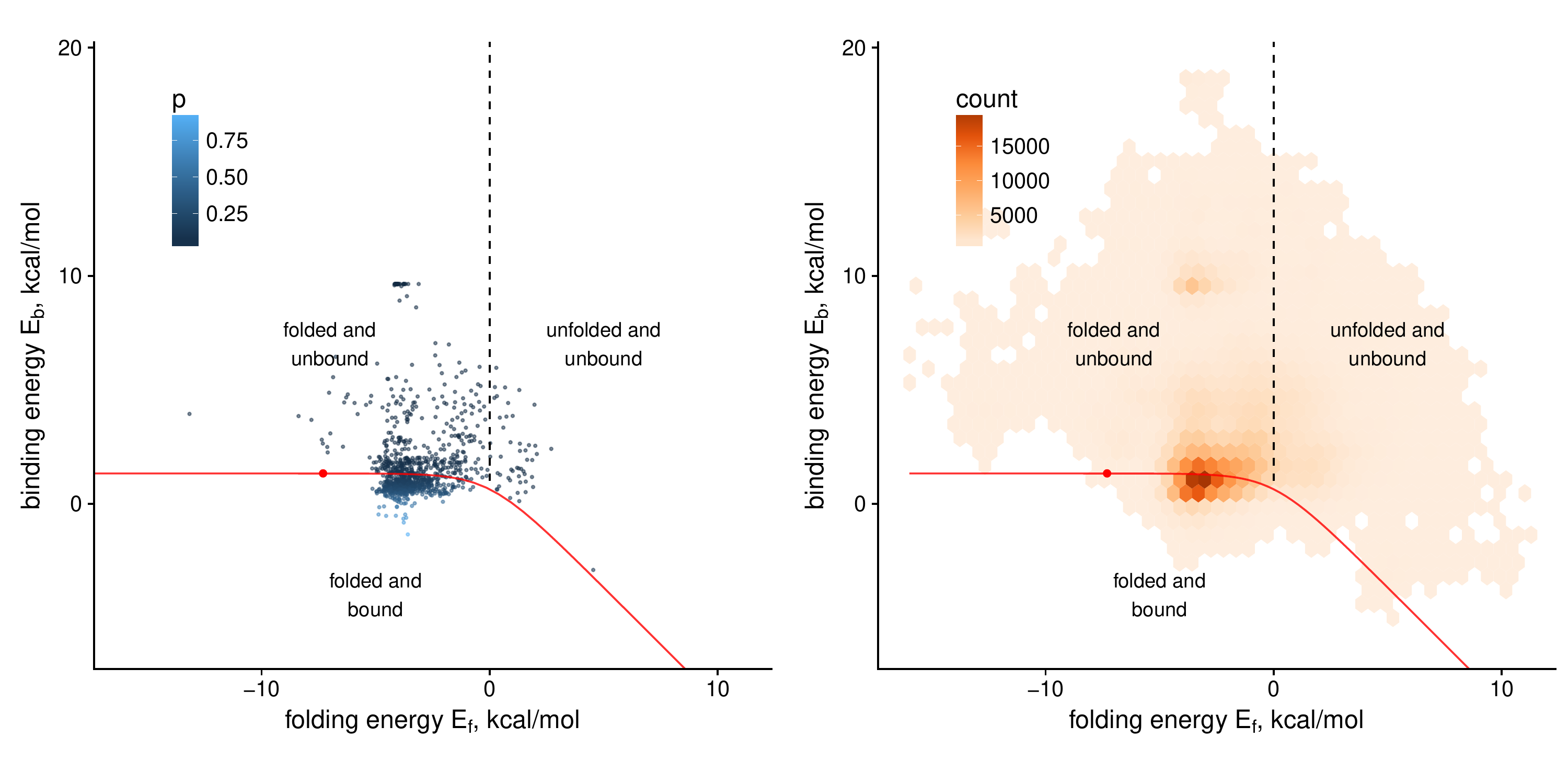}
\par\end{centering}
\caption{\label{fig:folding-vs-binding} Folding vs.~binding energy for single
(A) and double (B) mutants. Red dot is the wild-type energy, the red
line is where binding fraction is the same as wild-type $p=p^{W}$,
and below the red line variants have $p>p^{W}$. The three equilibrium
states are labeled, with $E_{f}=0$ demarcating the folded and unfolded
states.}
\end{figure}

To assess how much sampling noise influences our results, we calculate
95\% confidence intervals of $\epsilon_{f}$ and $\epsilon_{b}$ from
the bootstrapped estimates (Fig.~\ref{fig:bootstrap}), and find
that they are very narrow compared to the range of effect sizes for
most of the 2092 energy parameters. Examining $\epsilon_{f}$ and
$\epsilon_{b}$ across sites and amino acids (Fig.~\ref{fig:efiebi})
reveals a detailed picture of how folding and binding are sensitive
to substitutions. The energies have striking differences in their
patterns, and $\epsilon_{f}$ and $\epsilon_{b}$ are uncorrelated
(Fig.~\ref{fig:folding-vs-binding}A, $\rho=0.03$, $p_{value}=.28$).
Some substitutions have strong antagonistic effects, such as at positions
23 and 41, neither of which are at the binding interface. The substitutions
41L and 54G have particularly strong antagonism, although the double
mutant is known to be strongly epistatic, and there may be systematic
errors in these parameters from effects not captured by our model.
With relatively few exceptions, amino acid substitutions in GB1 do
not produce trade-offs between binding and fold stability. 

The energy landscape of a protein is determined by its structure,
so we expect that the inferred energies are related to structural
features. Both $\epsilon_{f}$ and $\epsilon_{b}$ correlate with
residue depth (Fig.~\ref{fig:prediction}B), and $\epsilon_{b}$
has a weaker relation to depth, except for a few sensitive shallow
residues. The three most sensitive sites to binding are at the interface
of the two proteins \cite{sauer-eriksson_crystal_1995,sloan_dissection_1999}
(plus signs in Fig.~\ref{fig:prediction}B), but many other sites
at the interface are not sensitive. 

A top down view of folding vs.~binding energy of single and double
mutants depicts how the variants fall into each of the three states.
In this phenotypic space, the wild-type is better than most of the
observed sequences, and in terms of binding fraction, it is in the
72nd and 85th percentile of the single and double substitutions respectively.
Most variants fall in the region of excess stability ($E_{f}\ll0$),
whereas binding energies are distributed around the wild-type, implying
that binding fraction is more sensitive to changes in binding energy
than folding energy for the majority of variants. This is consistent
with the lack of correlation between additive energies from the two
state model and independently measured folding energies (Fig.~\ref{fig:ddg1}),
as well as the lack of correlation between changes in fitness and
folding energies \cite{olson_comprehensive_2014}. 

\subsection*{Patterns of epistasis}

\begin{figure}
\begin{centering}
\includegraphics[scale=0.6]{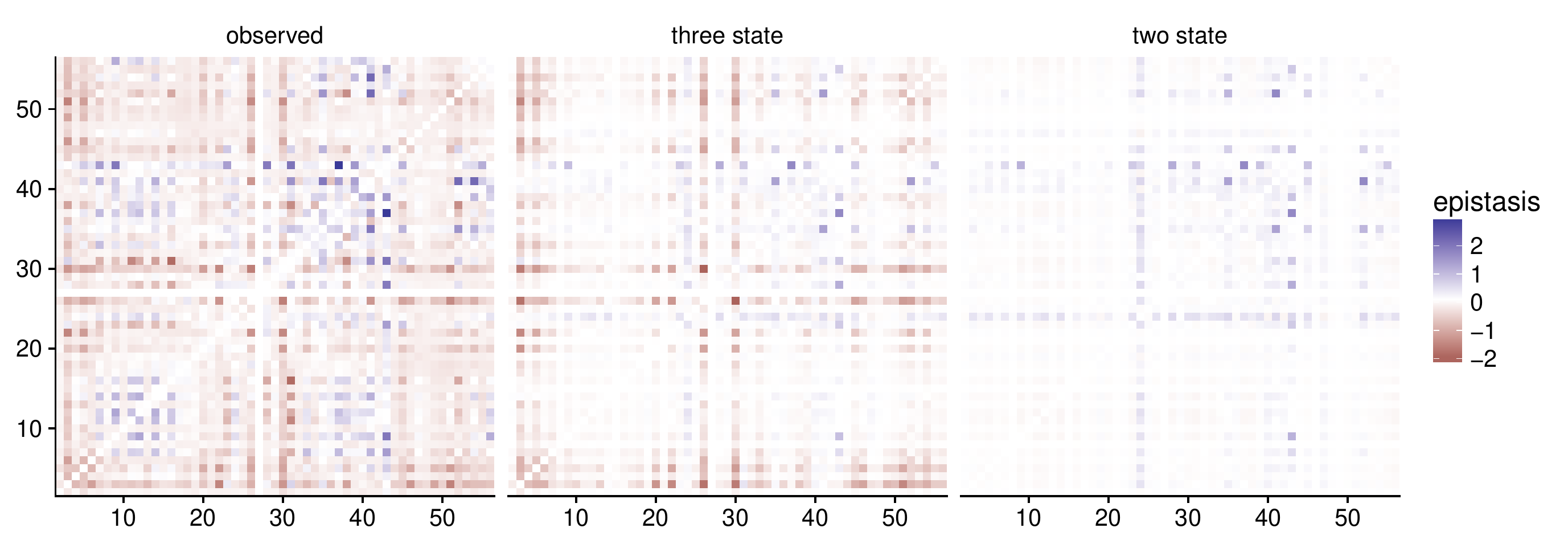}
\par\end{centering}
\caption{\label{fig:epistasis}Patterns of pairwise epistasis observed in the
data and predicted by the three state and two state thermodynamic
models. Shown are observed pairwise fitness epistasis (eq.~\ref{eq:epistasis}),
and inferred epistasis from the three state model and the two state
model. A network of residues that undergo correlated conformational
dynamics (positions 7, 9, 11, 12, 14, 16, 33, 37, 38, 40, 54, 56)
have significant observed positive epistasis that the thermodynamic
models fail to estimate. Epistasis is averaged for each pair of sites
over the relevant amino acid substitutions. We filter out non-biological
epistasis that is a consequence of the experimental limits on measured
fitness due to non-specific background binding (see Methods).}
\end{figure}

While the energies in our models are non-epistatic, the binding fraction
and fitness are epistatic due to the non-linearity between binding
fraction and energy (eq.~\ref{eq:p}), and our inferred pairwise
epistasis $\hat{J}_{ij}^{ab}$, analogous to eq.~\ref{eq:epistasis},
can be compared to the observed epistasis. We filter out non-biological
epistasis due to experimental limits on measured fitness, i.e., non-specific
background binding (see Methods), and average over amino acids in
each pair of sites. The predictions from the three-state model reproduces
the biological epistasis better than the two-state model, which vastly
underestimates the magnitude of epistasis across the protein (Fig.~\ref{fig:epistasis}).
Notably, the three state model predicts much of the negative epistasis,
but misses clusters of positive epistasis. 

To quantify these deviations, the difference between the predicted
and observed epistasis can be normalized by the noise in the observed
epistasis, $z_{ij}=\frac{\hat{J}_{ij}-J_{ij}}{\sqrt{v_{J_{ij}}}}$
(see Methods). The clusters of positive epistasis are more clearly
visible after filtering out all but the most underestimated epistasis
(bottom 5\% of $z_{ij}$, Fig.~\ref{fig:epi2}). As noted in \cite{olson_comprehensive_2014},
the residues in these positions had correlated conformational dynamics
in NMR and molecular dynamics studies (positions 7, 9, 11, 12, 14,
16, 33, 37, 38, 40, 54, 56, \cite{clore_amplitudes_2004,lange_molecular_2005,markwick_exploring_2007}).
This suggests that this unexplained epistasis is due to systematic
errors not accounted for in the model, such as epistasis in energy
or some alternative conformations, and therefore large prediction
errors can identify sites that should be studied in more detail. 
 
We also made predictions for a follow up study by Wu et al.~\cite{wu_adaptation_2016},
which targeted 4 highly epistatic sites in GB1, and assayed all combinations
of mutations, i.e.~$20^{4}$ variants. Most of the mutants with 3
or 4 mutations have very weak binding, and the fitness predictions
from the model trained on the Olson et.~al data can roughly predict
their functionality (Fig.~\ref{fig:wupredict}). Our model predicts
functional quadruple mutants with a true positive rate of 86\% and
a true negative rate of 95\% (defining functional as $f>-2.5$, Tab.~\ref{tab:wu}).
At the same time, the fitness is underestimated for many variants
much more than expected from measurement variability (Fig\@.~\ref{fig:wuz}),
suggesting that, in this more mutated data, some unaccounted for epistasis
is restoring binding or stability for approximately 20\% of variants. 

\section*{Discussion}

We have shown how with a few biophysical assumptions, i.e.,.~a small
number of thermodynamic states and additivity of energy, are sufficient
to extract a detailed folding and binding energy landscape of a protein.
Many DMS studies use \textit{in vitro} and\textit{ in vivo} selection
assays, and quantify the results with enrichment ratios, similar to
fitness in eq.~\ref{eq:fitness}. Most of these studies focus on
single substitutions from a wild-type, however fitness changes of
single substitutions confound changes in stability and binding. We
have shown how combining information from multiple sequences with
many mutations provide the constraints to separate these phenotypes
in a thermodynamic model, and therefore highly mutagenized sequence-function
experiments can provide a rich description of a protein's energy landscape. 

The inferred energies from the three state model very accurately predict
the independent low-throughput measurements, and show that it is feasible
to infer folding and binding energy in physical units accurately from
simple high-throughput functional assays. Several DMS studies have
indirectly measured stability. Araya et al.~\cite{araya_fundamental_2012}
applied a metric, based on the rescue effect of double mutations,
to identify stabilizing mutations, and Rocklin et al.~\cite{rocklin_global_2017}
used proteolysis assays that correlate with stability. Olson et al.~\cite{olson_comprehensive_2014}
extracted stability measures from this dataset by searching for single
mutation fitness changes in different genetic backgrounds that correlated
with the literature set of stability measurements. Further refinements
were developed with clustering methods that use structural information
and physiochemical properties \cite{wu_high-throughput_2015}. However,
these ad-hoc methods can suffer from over-fitting and require extraneous
knowledge to work effectively. In contrast, the thermodynamic model
directly infers stability in physical units and does not require any
benchmark stability measurements or structural information. 

Our method also infers the binding energies that are marginal to the
folded\textendash unbound state and measure the destabilization of
the binding interface. In constrast, dissociation constants, as measured
by titration curves, do not account for differences in fold stability.
The recently developed tite-seq method \cite{adams_measuring_2016}
infers a saturation constant to account for sequence dependent differences
in stability and expression, but does not compensate for stability
effects in the dissociation constant itself.

We have shown that the three state model shows good agreement with
patterns of epistasis, and deviations from our model identify a network
of residues that have correlated conformational dynamics. Since the
deviations and measurement noise itself can be rather large, sign
epistasis, path accessibility, and other geometric features of the
inferred genotype-phenotype map are likely to be distorted. It is
possible that more complex models, such energetic interaction terms
or more conformational states, can describe the remaining epistasis
in double mutants and in variants with more than two mutations. The
three state model works well for the relatively small GB1, but larger
proteins may need additional states, such as mis-folded conformations,
to accurately model their properties. 

Inferred energy landscapes from DMS may also be useful in understanding
protein structure. Inferred energy parameters may be useful for calibrating
potential functions used in structure prediction \cite{lee_ab_2017}.
Refined thermodynamic models with pairwise epistatic energy may be
able to infer protein contacts directly, similar to how multiple sequence
alignments of homologous proteins can infer contacts \cite{morcos_direct_2014},
providing a way to predict structure from DMS studies. Thermodynamic
models coupled with DMS also provide a way to study intrinsically
disordered proteins, which fold and bind simultaneously, but have
no persistent structure while unbound \cite{wright_linking_2009}.
Since many conformations can correspond to these states, free energy
differences are a natural way to quantify the properties of disordered
proteins.

While we have inferred a detailed genotype-phenotype map of GB1, yet
we do not know the consequences for evoltuion, which depend on how
the binding fraction affects the organismal fitness. Manhart and Morozov
\cite{manhart_protein_2015} explored the evolutionary dynamics of
a fitness function that is a linear combination of the three protein
states. This leads to an evolutionary coupling between  binding and
folding, where selection on folding can drive changes in binding,
and vice versa. With appropriate data it may be possible to infer
selection coefficients associated with each state, as well as evolutionary
trajectories in energy space, from multiple sequence alignments.

\paragraph*{}

\section*{Methods}

\subsection*{Poisson likelihood for an \textit{in vitro} selection assay}

In an \textit{in vitro} selection assay with one round, the library
of protein variants is sequenced before and after binding, and therefore
the count or multiplicity of each sequence carries information on
the binding. For each variant $\sigma$, with initial and final counts
$n_{0}$ and $n_{1}$, we define a Poisson log-likelihood with intensity
$\lambda_{0}=N_{0}$ and $\lambda_{1}=N_{0}pr$, where $p$ represents
the fraction of bound protein and $r$ is the systematic difference
between initial and final sequencing. We omit the dependence on $\sigma$
for brevity (note that $r$ does not depend on $\sigma$). The per
variant joint likelihood over the two time points is
\begin{equation}
LL=-N_{0}(1+pr)+n_{0}\log(N_{0})+n_{1}\log(N_{0}pr),
\end{equation}
omitting terms that don't depend on parameters. This has a nuisance
parameter $N_{0}$ per sequence, which has an ML estimate, given the
other parameters 
\[
N_{0}^{*}=\frac{n_{0}+n_{1}}{1+pr}
\]
Plugging it into the likelihood and dropping terms which don't depend
on the parameters results in
\begin{equation}
LL=-(n_{0}+n_{1})\log(1+pr)+n_{1}\log(pr).\label{eq:LL}
\end{equation}
The maximum likelihood estimate of binding fraction is $p^{\prime}=\frac{n_{1}}{n_{0}r}$.
Since some of the counts can be very small, we add a pseudo-count
of $\frac{1}{2}$ to $n_{0}$ and $n_{1}$ to slightly regularize
the estimate.

\subsection*{Thermodynamic model inference}

We use the likelihood in eq.~\ref{eq:LL}, and parameterize $p$
as a thermodynamic model following the Boltzmann distribution. We
modify $p$ to account for non-specific background binding $p_{0}$:
\begin{equation}
p(\sigma)=\frac{1}{1+e^{E_{b}(\sigma)}(1+e^{E_{f}(\sigma)})}(1-p_{0})+p_{0},
\end{equation}
and the energies have an additive relation to sequence defined by
eqs.~\ref{eq:ef}\ref{eq:eb}. The total likelihood is the sum of
the per variant likelihoods $\textbf{LL}=\sum_{\sigma}LL(\sigma)$.

This log-likelihood is non-convex, and is optimized using the NLopt
library \cite{steven_g._johnson_nlopt_nodate}, which implements the
method of moving asymptotes algorithm \cite{svanberg_class_nodate},
and uses the log-likelihood gradients with respect to $r$, $p_{0}$,
and the energies. The initial parameters were $r=1$, $p_{0}=0.01$,
and all energies set to zero. $r$ and $p_{0}$ were reparameterized
as $e^{r^{\prime}}$ and $e^{p_{0}^{\prime}}$ inside the optimization
function, so that the original parameters are non-negative. In the
optimization algorithm, upper and lower bounds on $\varepsilon$ are
set to limit very small gradients which stop the optimization prematurely.
The value of to $\pm15$ was chosen by optimizing with different bounds,
$\pm10$, $\pm15$, $\pm20$, and $\pm25$, and choosing the result
with the highest likelihood. Dimensionless energies are converted
to kcal/mol with $T=297$, and therefore the bounds are $\pm8.85$
kcal/mol.

\subsection*{Bootstrap}

Since the optimization algorithm can get stuck in local optima, we
use a bootstrap restarting procedure to overcome local optima related
to sampling noise \cite{wood_minimizing_2001}, and to generate a
bootstrap distribution of parameters to quantify their uncertainty
due to sampling noise. The maximum likelihood parameters from the
fit described above, $\theta$, are the initial parameters in an iterative
procedure that alternates optimizing on the original and bootstrapped
data. 

Each iteration consists of: 1) creation of bootstrapped data with
counts drawn from a Poisson distribution with means $n_{0}$ and $n_{1}$.
2) Searching for the maximum likelihood parameters $\theta^{\prime}$
on the bootstrapped data with initial parameters $\theta$. 3) Searching
for the maximum likelihood parameters $\theta^{\prime\prime}$ on
the original data with initial parameters $\theta^{\prime}$. 4) If
the likelihood is no better than the best optimization within a small
threshold $LL(\theta^{\prime\prime})\le LL(\theta)+\eta$ ($\eta=0.0001$
), then add the bootstrapped parameters $\theta^{\prime}$ to a list.
5) If the likelihood is better than the previous best $LL(\theta^{\prime\prime})>LL(\theta)+\eta$,
then update the best parameters, $\theta\leftarrow\theta^{\prime\prime}$,
and delete the list of bootstrapped parameters. 6) terminate the procedure
once 100 bootstraps have been accumulated in the list.

\subsection*{Pairwise epistasis}

A minimal amount of non-specific background binding, $p_{0}$, imposes
a lower bound on measured binding fraction in the experiment, estimated
to be $f_{0}=\log(p_{0}/p(\sigma^{W})=-5.69$ by our three state model.
This threshold effect produces large amounts of non-biological positive
pairwise epistasis, e.g.~when the double mutant has the same level
of binding as one of the single mutants at the background level. Therefore,
the data shown in Fig.~\ref{fig:epistasis} excludes $\hat{J}_{ij}^{ab}$
with sequences near this threshold, i.e., $\min\left(f(\sigma_{/(i,a)/(j,b)}^{W}),f(\sigma_{/(i,a)}^{W}),f(\sigma_{/(j,b)}^{W})\right)<-4.5$. 

\subsection*{Sample variance of fitness and epistasis}

Since estimated fitness is asymptotically Gaussian, the sample variance
of fitness is equal to the curvature of the log-likelihood surface.
Replacing $p^{\prime}$ with $e^{f^{\prime}}$ in eq.~\ref{eq:LL},
the asymptotic variance of $f^{\prime}$ is $-\left(\frac{\partial^{2}LL}{\partial f^{\prime2}}\right)^{-1}$.
The variance of the fitness estimate, as defined in eq.~\ref{eq:fitness}
is the sum of the focal and wild-type variances
\begin{equation}
v_{f}(\sigma)=\frac{n_{0}(\sigma)+n_{1}(\sigma)}{n_{0}(\sigma)n_{1}(\sigma)}+\frac{n_{0}(\sigma^{W})+n_{1}(\sigma^{W})}{n_{0}(\sigma^{W})n_{1}(\sigma^{W})},
\end{equation}
and the variance in epistasis is the sum of the single and double
mutant variances
\[
v_{J_{ij}^{ab}}=v_{f}(\sigma_{/(i,a)/(j,b)}^{W})+v_{f}(\sigma_{/(i,a)}^{W})+v_{f}(\sigma_{/(j,b)}^{W}),
\]
The variance is then averaged over amino acids $a,b$ for each position
$i,j$.

\appendix
\newpage{}

\section*{Supplemental Material}

\renewcommand{\thefigure}{S\arabic{figure}}
\setcounter{figure}{0}
\renewcommand{\thetable}{S\arabic{table}}
\setcounter{table}{0}

\pagebreak{}

\begin{figure}
\begin{centering}
\includegraphics[scale=0.6]{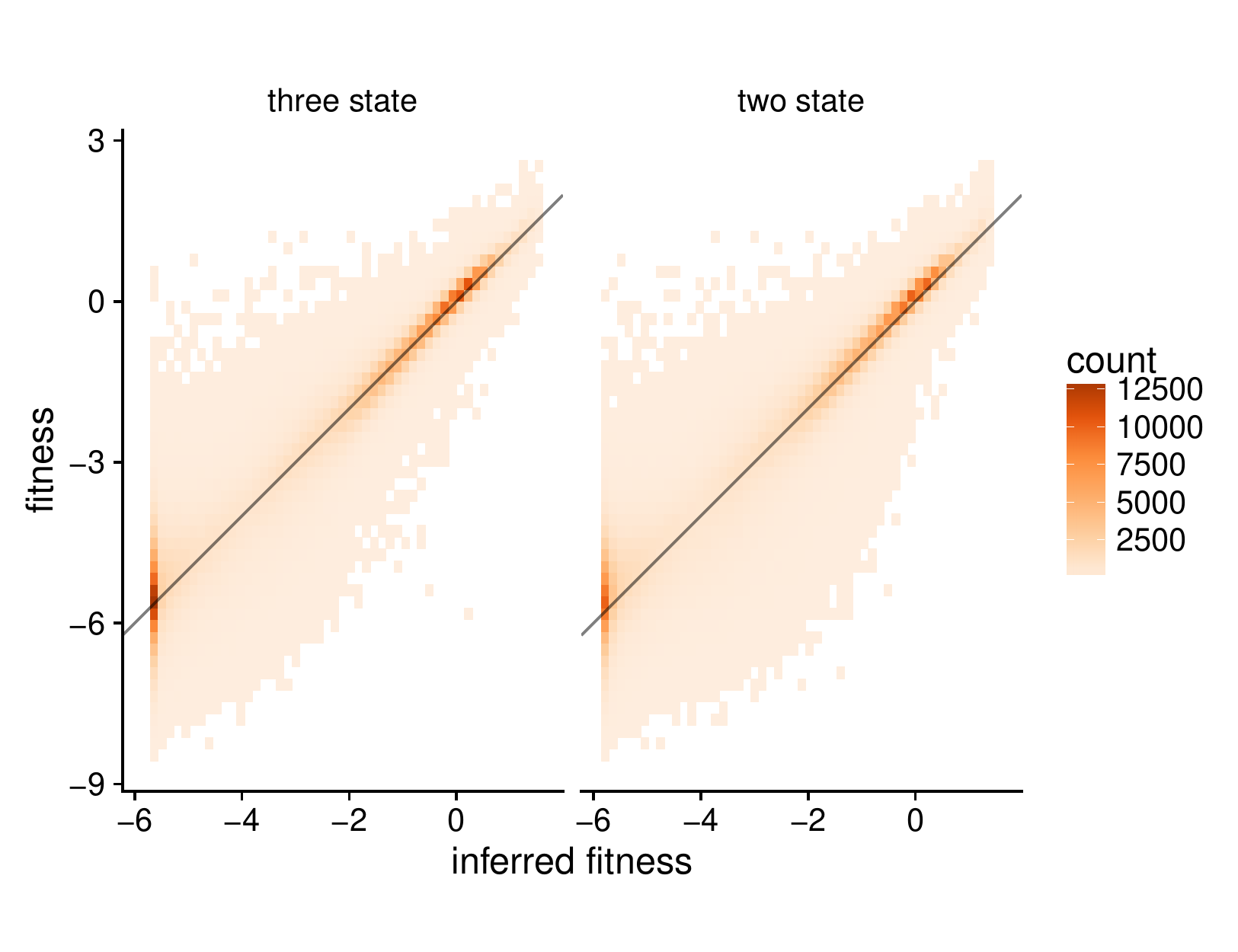}
\par\end{centering}
\caption{\label{fig:yyhat} Inferred versus predicted fitness for two state
model and three state model. Correlations are 96.4\% and 97.1\% respectively.}
\end{figure}
\begin{figure}
\begin{centering}
\includegraphics[scale=0.6]{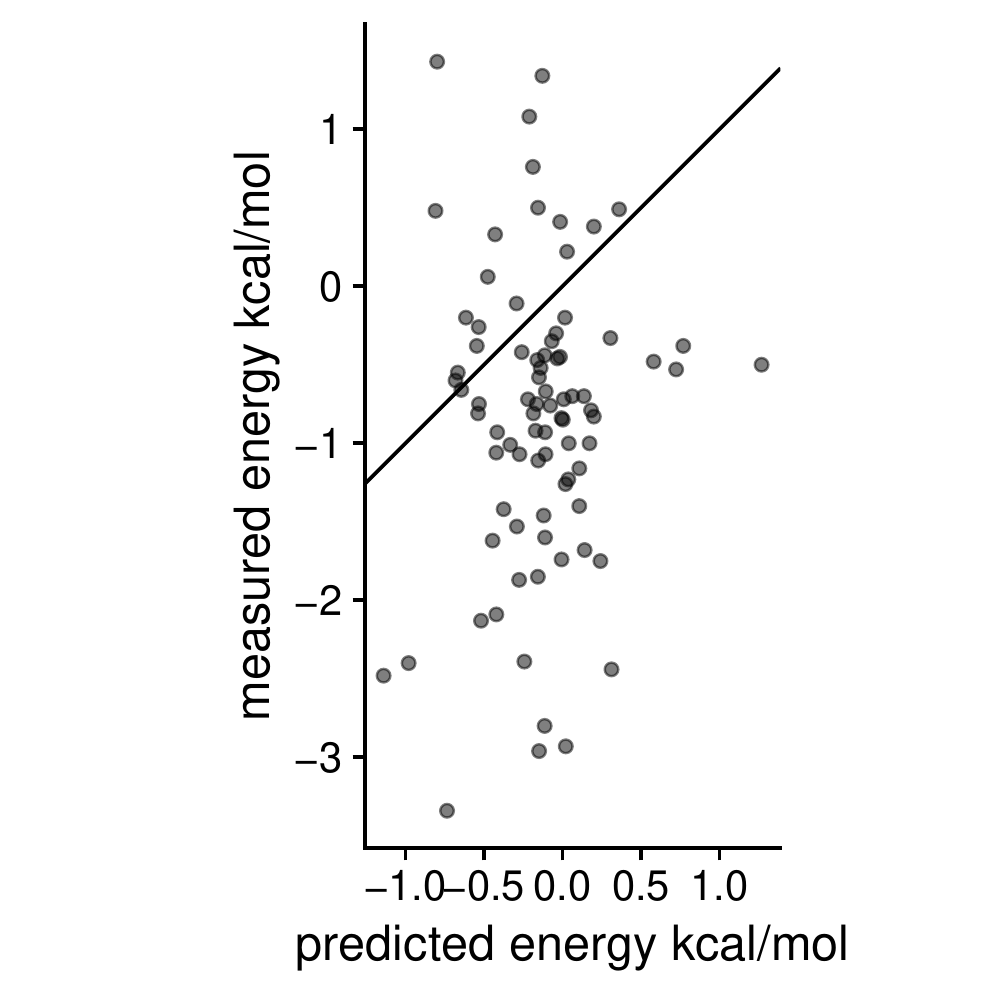}
\par\end{centering}
\caption{\label{fig:ddg1}Additive energies from two state model do not predict
changes in fold stability. Measured $\epsilon$ same as in fig.~\ref{fig:prediction}B.
If variants have excess stability, the measured binding fraction would
be mostly sensitive to binding, which would lead to energies unrelated
to folding.}

\end{figure}
\begin{figure}
\begin{centering}
\includegraphics[scale=0.6]{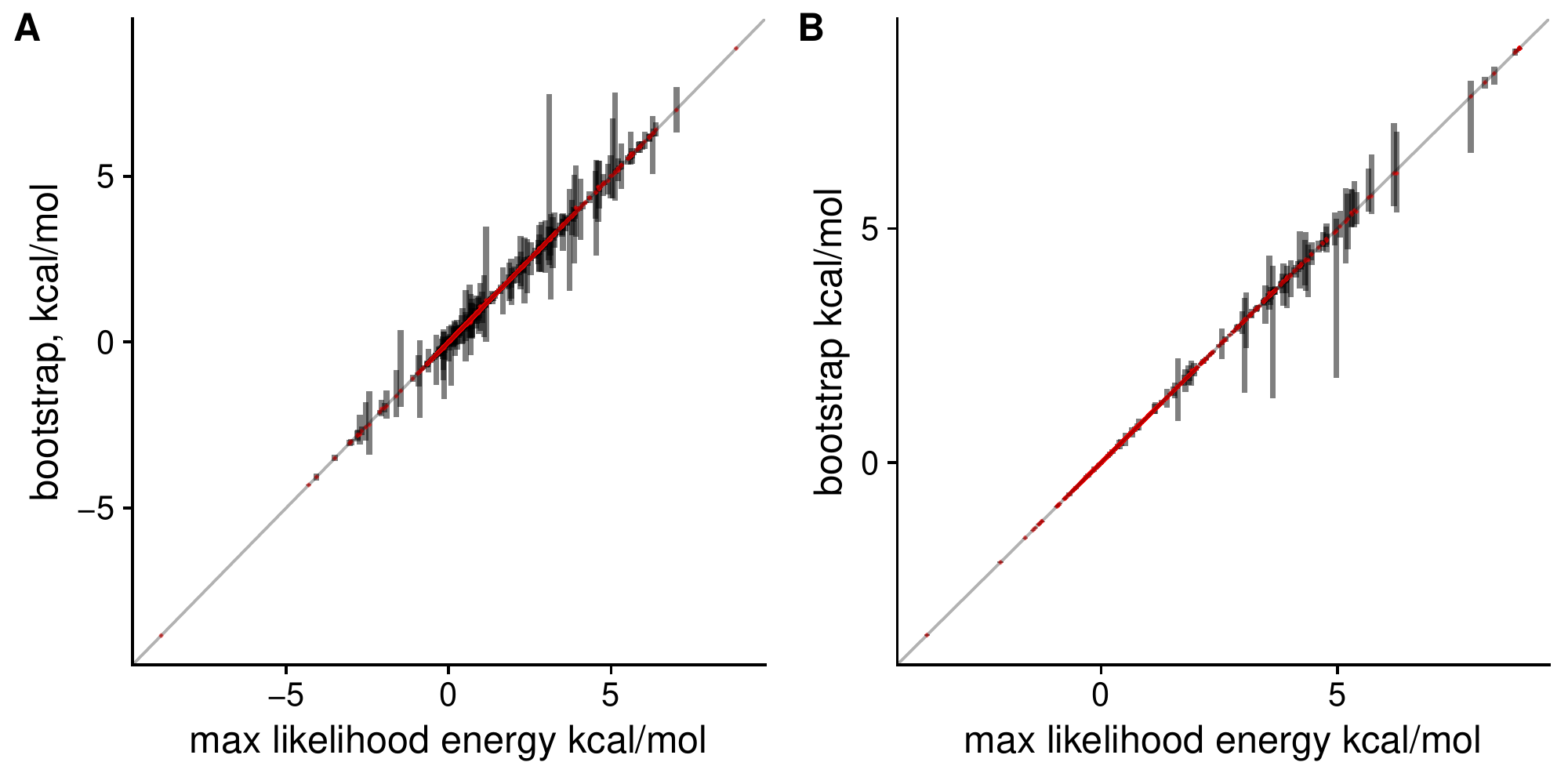}
\par\end{centering}
\caption{\label{fig:bootstrap}Bootstrapped parameters of the three state model,
for the folding (A) and binding (B) energies. Red points denote the
median value from the bootstrap, and the gray bars show the 95\% confidence
interval.}

\end{figure}
\begin{figure}
\begin{centering}
\includegraphics[scale=0.6]{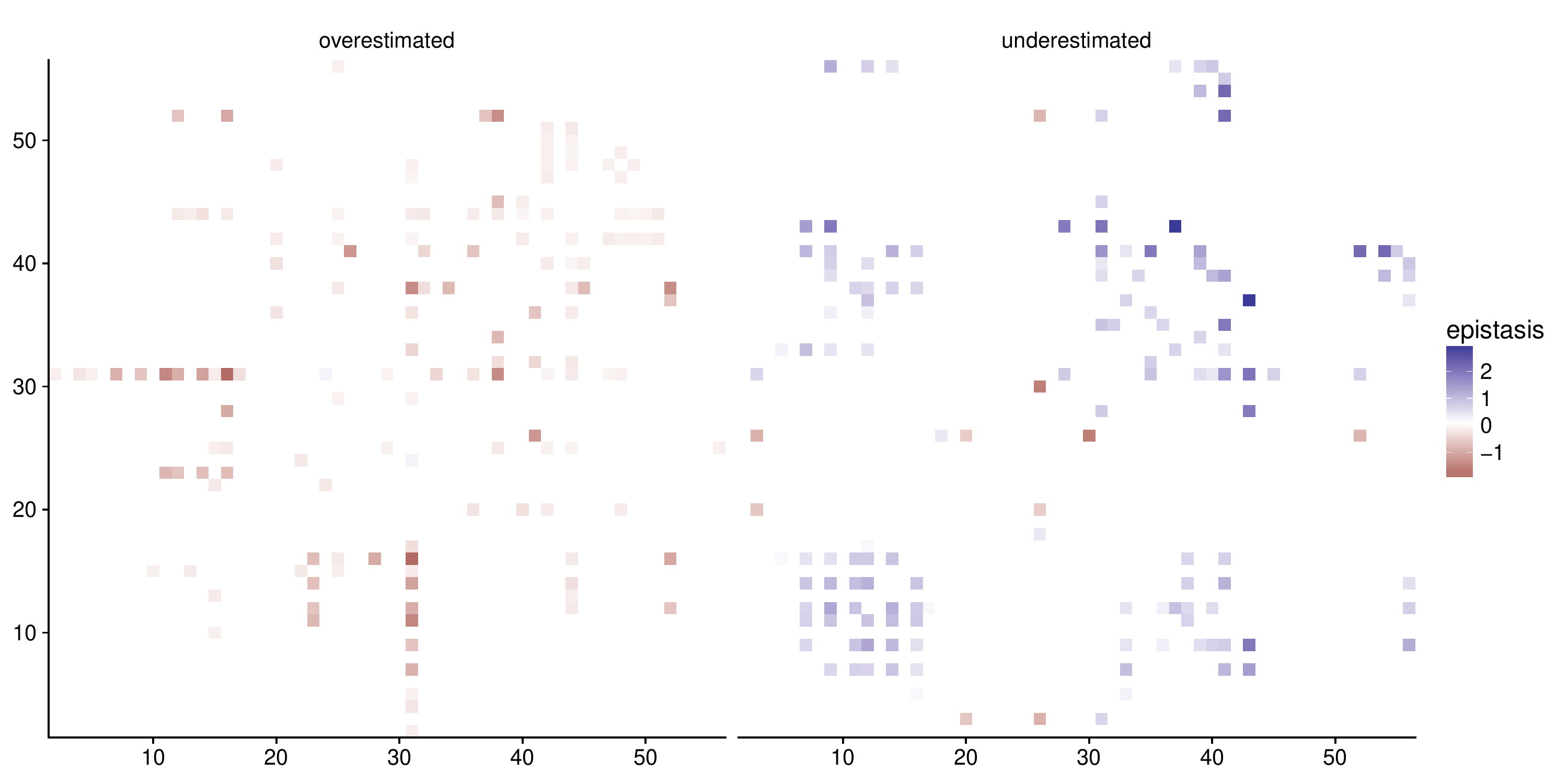}
\par\end{centering}
\caption{\label{fig:epi2} Observed epistasis $J_{ij}$ (averaged over amino
acids) which our model overestimates ($\hat{J}_{ij}\gg J_{ij}$) and
underestimates ($\hat{J}_{ij}\ll J_{ij}$). Overestimated epistasis
is the top 5\% of $z_{ij}$ and underestimated epistasis is the bottom
5\%. Underestimated epistasis is largely positive and corresponds
to the dynamically correlated residue network.}

\end{figure}
 
\begin{figure}
\begin{centering}
\includegraphics[scale=0.6]{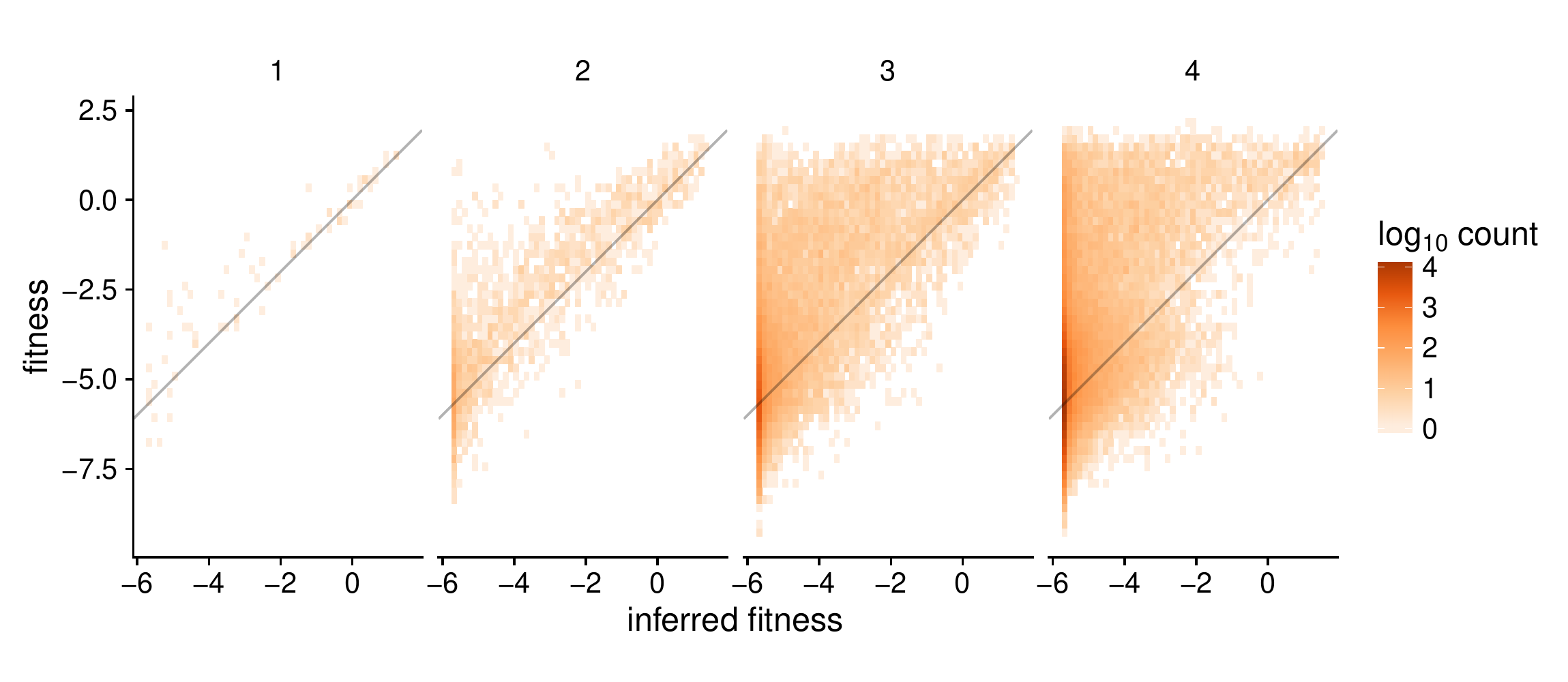}
\par\end{centering}
\caption{\label{fig:wupredict}Fitness is less predictable in a follow-up study
that targeted all combinations at four sites \cite{wu_adaptation_2016}.
Panels show true and inferred fitness for 1 to 4 substitutions from
wild-type. A substantial fraction of functional variants are underestimated,
suggesting some unaccounted for epistasis in energy or conformational
dynamics. See also Fig.~\ref{fig:wuz} and Tab\@.~\ref{tab:wu}.}
\end{figure}
 
\begin{figure}
\begin{centering}
\includegraphics[scale=0.6]{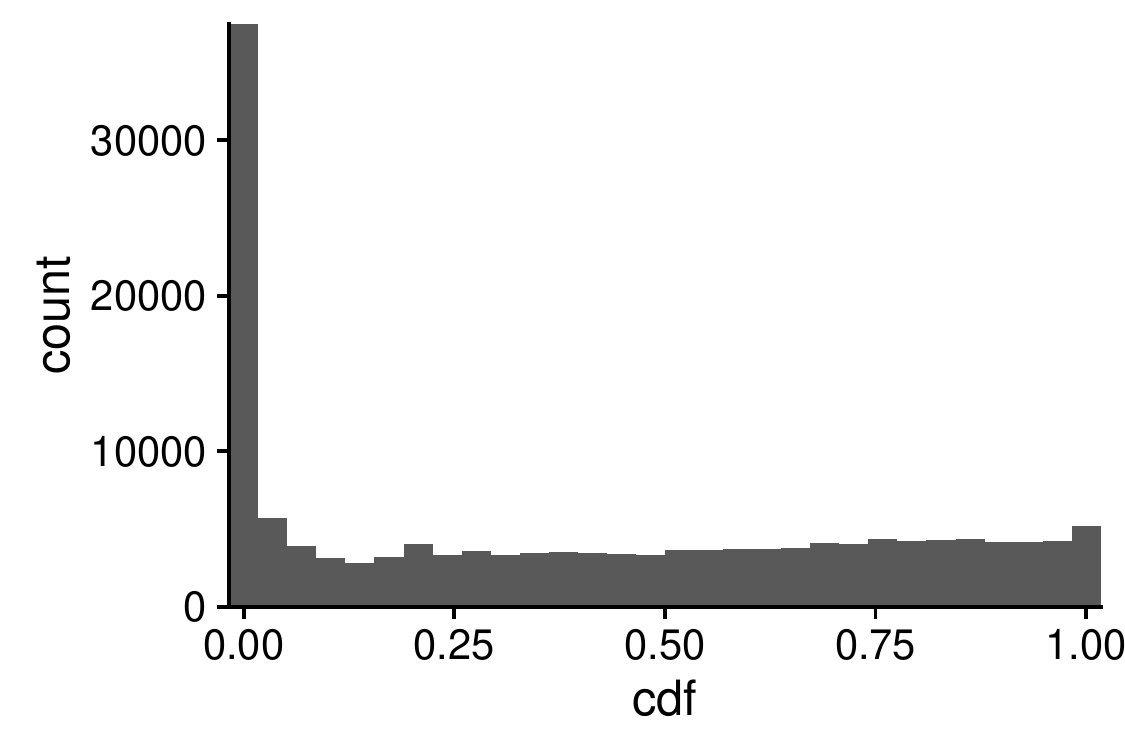}
\par\end{centering}
\caption{\label{fig:wuz}Around 20\% of variants in Wu et al predictions have
fitnesses underestimated more than expected from measurement variation.
Shown is distribution of $\textrm{CDF}\left(\frac{\hat{f}(\sigma)-f(\sigma)}{\sqrt{v_{f}(\sigma)}}\right)$,
where $\textrm{CDF}$ is the cumulative distribution function of a
standard normal distribution.}

\end{figure}
 
\begin{table}
\begin{centering}
\begin{tabular}{|c|c|c|c|c|c|c|}
\hline 
$d_{h}$ & $n$ & $\rho$ & TP & FP & TN & FN\tabularnewline
\hline 
\hline 
1 & 76 & 0.95 & 39 & 1 & 29 & 7\tabularnewline
\hline 
2 & 2091 & 0.74 & 498 & 19 & 1342 & 232\tabularnewline
\hline 
3 & 26019 & 0.42 & 1596 & 158 & 21355 & 2910\tabularnewline
\hline 
4 & 121174 & 0.39 & 982 & 162 & 113984 & 6046\tabularnewline
\hline 
\end{tabular}
\par\end{centering}
\caption{\label{tab:wu}Statistics for predictions on Wu et.~al data. $d_{h}$:
hamming distance from wild-type. $n$: number of variants. $\rho$:
correlation coefficient weighted by $1/v_{f}$. TP, FP, TN, FN: True/false
positives/negatives for whether the variant is functional ($f>-2.5$). }

\end{table}

\end{document}